\documentclass[twocolumn,superscriptaddress,aip,jcp]{revtex4}

\usepackage{amsmath,amssymb,amsfonts,bm}
\usepackage{graphicx,epstopdf}  %epsfig,
\usepackage{color}
\usepackage{lscape}
\usepackage{extarrows}
\usepackage{enumerate}
\usepackage{empheq}
\usepackage{multirow}

\usepackage{array}
\usepackage{booktabs}

\newcommand{\cm}{\mathrm{cm}^{\tiny \text{--}1}}

\renewcommand{\d}{{\rm d}}

\newcommand{\M}{\mbox{\tiny M}}

\newcommand{\w}{\omega}
\newcommand{\wti}{\widetilde}
\newcommand{\ti}{\tilde}

\newcommand{\B}{\mbox{\tiny B}}

\newcommand{\s}{\mbox{\tiny S}}

\newcommand{\tS}{\mbox{\tiny S}}

\newcommand{\T}{\mbox{\tiny T}}

\newcommand{\env}{\mbox{\tiny env}}
\newcommand{\exc}{\mbox{\tiny ex}}

\newcommand{\nul}{\mbox{\tiny vib}}

\newcommand{\dg}{\dagger}
\newcommand{\la}{\langle}
\newcommand{\ra}{\rangle}

\newcommand{\Sec}[1]{Sec.\,\ref{#1}}
\newcommand{\App}[1]{Appendix \ref{#1}}
\newcommand{\nl}{\nonumber \\}
\newcommand{\be}{\begin{equation}}
\newcommand{\ee}{\end{equation}}
\newcommand{\bsube}{\begin{subequations}}
\newcommand{\esube}{\end{subequations}}
\newcommand{\Eq}[1]{Eq.\,(\ref{#1})}
\newcommand{\Eqs}[1]{Eqs.\,(\ref{#1})}
\newcommand{\Fig}[1]{Fig.\,\ref{#1}}

\newcommand{\RN}[1]{%
  \textup{\uppercase\expandafter{\romannumeral#1}}%
}

\begin{document}
\title{
Correlated vibration--solvent effects on
the non-Condon exciton spectroscopy
}

\author{Zi-Hao Chen}
\affiliation{{Department of Chemical Physics, University of Science and Technology of China, Hefei, Anhui 230026, China}}

\author{Yao Wang}
\email{wy2010@ustc.edu.cn}
\affiliation{Hefei National Laboratory for Physical Sciences at the Microscale
 and iChEM and Synergetic Innovation Center of Quantum Information
 and Quantum Physics,
 University of Science and Technology of China, Hefei, Anhui 230026, China}

\author{Rui-Xue Xu}
\email{rxxu@ustc.edu.cn}
\affiliation{{Department of Chemical Physics, University of Science and Technology of China, Hefei, Anhui 230026, China}}
\affiliation{Hefei National Laboratory for Physical Sciences at the Microscale
and iChEM and Synergetic Innovation Center of Quantum Information
and Quantum Physics,
University of Science and Technology of China, Hefei, Anhui 230026, China}

\author{YiJing Yan}
\affiliation{{Department of Chemical Physics, University of Science and Technology of China, Hefei, Anhui 230026, China}}
\affiliation{Hefei National Laboratory for Physical Sciences at the Microscale
 and iChEM and Synergetic Innovation Center of Quantum Information
 and Quantum Physics,
 University of Science and Technology of China, Hefei, Anhui 230026, China}

\date{\today}
\begin{abstract}
Excitation energy transfer is crucially involved in
a variety of systems. During the process, the non-Condon vibronic coupling
and the surrounding solvent interaction may synergetically play important roles.
In this work, we study the correlated vibration--solvent influences on the non-Condon
exciton spectroscopy.
Statistical analysis is elaborated
for the overall vibration--plus--solvent environmental effects.
Analytic solutions are derived for the linear absorption of monomer systems.
General simulations are accurately carried out
via the dissipaton--equation--of--motion approach.
The resulted spectra in either the linear absorption or strong field regime
clearly demonstrate the coherence enhancement due to the synergetic vibration--solvent correlation.
\end{abstract}
\maketitle

\section{Introduction}
\label{sec1}

  Excitation energy transfer (EET) is crucially involved in
a variety of systems including molecular aggregates, organic semiconductors, light harvesting systems, etc.%
\cite{Sch06683,Hes187069,For19025005,Gin201,Nel202215}
Particularly, the long--lived quantum beats in photosynthetic antenna complexes have
aroused great interest.\cite{Eng07782,Che09241,Pan1012766,Ple13235102}
It is found that
the non-Condon vibronic coupling would affect
this quantum coherence enhancement.\cite{Kam0010637, Ish082219, Wom12154016}
Spectroscopic studies on EET systems have been widely carried out to reveal the structures and
underlying dynamic processes in these systems.\cite{Tur111904,Tem1412865,But14034306,Cam1595,Bal1519491,Zha16204109}

In reality, EET constitutes an open quantum process where intramolecular
nuclear vibrations and
the surrounding solvent would coordinatively play important roles.
To study the non-Condon vibronic coupling, one
needs to deal with the hybridization between the excitonic system
and nuclear vibrations.
By the Herzberg--Teller approximation,
vibrations would contribute to
the transition dipole moments. While the solvent, coupled to both excitons and vibrations, is considered to be non-polarizable.

Take, for example, a monomer excitonic system. The total Hamiltonian in the presence of external field reads
\bsube\label{Htot0}
\begin{align}
 H_{\T}(t) = H_{\M}-\varepsilon(t)\hat \mu_{\T},
\end{align}
where
\begin{align}\label{HMtot}
 H_{\M} = (\Delta+H'_{\env}-H_{\env})\hat{B}^{\dg}\hat{B}+H_{\env}\,.
%-\varepsilon(t)\hat \mu_{\T}.
\end{align}
\esube
Here, $\hat{B}\equiv |0\ra\la 1|$ ($\hat{B}^{\dag}\equiv |1\ra\la 0|$) is the excitonic annihilation (creation) operators, and $\Delta$ is the excitation energy. The total electronic--plus--vibrational dipole moment $\hat \mu_{\T}$ interacts with a classical external field $\varepsilon(t)$. $H_{\env}$ and $H'_{\env}$ are the vibration--plus--solvent environment Hamiltonians,  associated with the ground and excited states, respectively.
In this work, we adopt the Caldeira--Leggett's model,\cite{Cal83587,Wei12,Yan05187} i.e.\  
%%%
\begin{align}
\label{Henv}
 H_{\env}
&=\frac{\Omega}{2}(\hat p^2+\hat q^2)
+\sum_k\frac{\w_k}{2}
 \Big[\hat p^2_k+\big(\hat x_k-\frac{c_k}{\w_k}\hat q\big)^2\Big].
%+H_{\nul\text{-}\B}.
\end{align}
It involves a vibrational mode $\hat q$ and its coupling to
a solvent bath,
$h_{\B}=\frac{1}{2}\sum_k\w_k(\hat p^2_k+\hat x^2_k)$.
 $H'_{\env}$
is similar to $H_{\env}$ of \Eq{Henv}, but with linearly displaced $q'=q-D$ and $\{x'_k=x_k-d_k\}$. This results in an overall reorganization,
\begin{align}\label{lambda_def}
   \lambda&\equiv\la H'_{\env}-H_{\env}\ra_{\env}.
\end{align}
Here, $\la \hat O\ra_{\env}
  \equiv {\rm Tr}_{\env}(\hat O e^{-\beta H_{\env}})
/{\rm Tr}_{\env}e^{-\beta H_{\env}}$.
The environment induced force is given by
%%%
\begin{align}\label{force}
  \hat F=H'_{\env}-H_{\env}-\lambda.
\end{align}

In this work, we are interested in the statistically correlated vibration--and--solvent influences onto the
EET systems and the associated non-Condon effect in the Herzberg--Teller approximation.
One approach could be inclusion of the vibrational mode into the system, leaving the solvent alone as
environment. This will enlarge the dimension of the system,
making it difficult to extend to multi-mode cases and polymeric systems.
The alternative approach is to treat the intramolecular vibrational mode and the solvent
altogether as environment, interacting with the EET system,
 in a statistical manner. This work starts from this strategy.
 However, most traditional quantum dissipation theories,
for example, the Redfield equation and its modified series,\cite{Red651,Lin76119,Gor76821,Her9111586,Yan982721,Yan002068,%
Nov0410363,Sch06084903} focus explicitly only on the reduced system,
making them practically intractable to simulate the environment--polarized excitation.
The exact hierarchical--equation--of--motion (HEOM) formalism in Gaussian environments,\cite{Tan89101,Yan04216,Ish053131,Tan06082001,Xu05041103,Xu07031107,%
Jin08234703,Che11194508,Ye16608,Tan20020901}
constructed via the calculus on path integrals, are composed of coupled differential equations
between the primary reduced system density operator and a set of auxiliary operators.
These auxiliary operators, although in principle, carry the environmental informations,
their associated quantum dynamic algebras make the HEOM formalism not convenient in the extension to study environment polarizations.

The dissipaton equation of motion(DEOM),\cite{Yan14054105,Zha15024112,Yan16110306,Xu18114103}
using quasi-particle descriptions for environments,
provides a unified treatment on hybridized environment dynamics
and entangled system--environment excitations, straightforwardly.
Being an exact method for Gaussian environment,
the DEOM recovers the HEOM
for just the reduced system dynamics.
Besides, the DEOM adopts ``dissipatons'' as quasi-particles
associated with the interacting environment statistical dynamics.\cite{Yan14054105,Zha15024112,Yan16110306,Xu18114103,Zha18780,Wan20041102}
It was applied to study the Herzberg--Teller vibronic coupling
in our previous work,\cite{Zha16204109} but the correlated solvent effects were not considered there.

This work starts from the physical model of total Hamiltonian, \Eq{Henv},
which contains the hybridized exciton--vibration, exciton--solvent, and vibration--solvent couplings.
The overall vibration--plus--solvent environment response function,
characterizing the environment statistical properties,
is briefed in \Sec{thsec2}, with the derivation detailed in \App{appa}.
A general DEOM formalism for multi-mode system--environment couplings and the Herzberg--Teller type of dipole moments
are constructed in \Sec{thsec3}.
Numerical demonstrations and discussions are presented in \Sec{thsec4}.
The analytic solution for the linear absorption of the current monomer system, \Eq{Henv}, is derived in \App{appb}.
The expression and analysis in the gas--phase limit is given in \App{appc}.
The paper is summarized in \Sec{thsec5}.

\section{Environment statistics}
\label{thsec2}

This section concentrates on the environment statistics.
The total system involves the correlated exciton--vibration, exciton--solvent, and
vibration--solvent couplings (cf.\ \Sec{sec2a})
where the vibrational mode in solvent behaves as a Brownian oscillator.
Following some elementary quantum statistical algebra,
the overall vibration--plus--solvent environment statistical properties,
basically the response functions, can be obtained (finalized in \Sec{sec2b} and detailed in \App{appa}).

\subsection{Vibration--solvent decomposition}
\label{sec2a}

According to \Eqs{Henv} and (\ref{lambda_def}), the overall reorganization energy reads
\begin{align}\label{lambda}
   \lambda=\frac{1}{2}\Omega D^2+\frac{1}{2}\sum_k
\w_k\Big(\frac{c_k}{\w_k}D-d_k \Big)^2         .
\end{align}
The environment induced force, defined in \Eq{force}, can be decomposed into two parts as
%%%
\begin{align}\label{F12}
  \hat F
   =\hat F_{1}+\hat F_{2},
\end{align}
where
\bsube\label{Fdecom}
\begin{align}
\label{Fvib}
   \hat F_{1}&=\Big[\sum_k c_kd_k-\big(\Omega+\sum_k\frac{c_k^2}{\w_k}\big)D\Big]\hat q,
\\
   \hat F_{2}&=D\hat X_{\B}-\hat Y_{\B},\label{FBXY}
\end{align}
\esube
with
\begin{align}\label{XBYB}
   \hat X_{\B}=\sum_kc_k\hat x_k\quad\text{and}\quad
   \hat Y_{\B}=\sum_k\w_kd_k\hat x_k.
\end{align}
Here $\hat F_1$ and $\hat F_2$ associate with the vibrational mode
and solvent degrees of freedom, respectively.
In this paper the former is considered of optical polarization
while the latter not.

It is noticed that the above environment model, \Eqs{Henv} and (\ref{Fdecom}),
constitutes a Gaussian--Wick's bath.  Its influence onto the system and the entangled system--environment dynamics can by totally characterized by the environment correlation functions
\begin{align}
 C_{ab}(t)\equiv \la \hat F_{a}^{\env}(t) \hat F_{b}^{\env}(0)\ra_{\env};
 \quad a,b=1,2.
\end{align}
Here, we have denoted
\be \label{ope_env}
 \hat O^{\env}(t) \equiv e^{iH_{\env}t}\hat O e^{-iH_{\env}t}.
\ee
Throughout this work, we set $\hbar=1$ and $\beta=1/(k_BT)$,
with $k_B$ being the Boltzmann constant and $T$ the temperature.
The correlations $C_{ab}(t)$ can be obtained via the fluctuation--dissipation theorem as\cite{Wei12,Yan05187}
\be\label{FDT}
C_{ab}(t)
=\frac{1}{\pi}\int^{\infty}_{-\infty}\!\!
  \d\w \frac{e^{-i\w t}  J_{ab}(\w)}{1-e^{-\beta\w}}.
\ee
In \Eq{FDT}, $J_{ab}(\w)$ are the interaction spectral densities, reading
\be\label{Jphi}
  J_{ab}(\w)=\frac{1}{2i}\int^{\infty}_{-\infty}\!\!\d t\, e^{i\w t} \Phi_{ab}(t),
\ee
where $\Phi_{ab}(t)$ are bath response functions defined as
\be\label{Phidef_sec2}
 \Phi_{ab}(t)\equiv i\la [\hat F_a^{\env}(t), \hat F_b^{\env}(0)]\ra_{\env}.
\ee
In the following parts of paper, we denote
    for any function $f(t)$ the frequency resolution,
\be\label{laplace}
   \ti f(\w)\equiv\int^{\infty}_{0}\!\!\d t\, e^{i\w t} f(t).  %\equiv \ti f^{(r)}(\w)+i\ti f^{(i)}(\w).
\ee

\subsection{Environment response functions}
\label{sec2b}

To obtain $J_{ab}(\w)$,  consider first the vibrational response function
in the total environment composite [\Eq{Henv}]
\be\label{chiqqt}
   \chi_{qq}(t) \equiv i\la[\hat q^{\env}(t), \hat q^{\env}(0)]\ra_{\env}.
\ee
Its frequency resolution is identified to be\cite{Wei12,Yan05187}
\be\label{chiqqw}
   \wti\chi_{qq}(\w)
    =\frac{\Omega}{\w^2_{\s}-\w^2-i\w\ti\zeta(\w)}.
\ee
The solvent induced frictional function, $\ti\zeta(\w)$,
is related to the solvent susceptibility function via\cite{Yan05187}
\be\label{iwzeta}
  i\w\ti\zeta(\w)=\Omega[\ti\varphi_{xx}(\w)-\ti\varphi_{xx}(0)].
\ee
In \Eq{iwzeta}, the involving interacting solvent response function is
\be\label{phixx}
  \varphi_{xx}(t)\equiv i\la[\hat X^{\B}_{\B}(t),
     \hat X_{\B}]\ra_{\B},
\ee
with
$\hat X_{\B}$ defined in \Eq{XBYB}.
Here, we denote
$
\la \hat O\ra_{\B}
  \equiv {\rm tr}_{\B}(\hat O e^{-\beta h_{\B}})
/{\rm tr}_{\B}e^{-\beta h_{\B}}
$
and
$\hat O^{\B}(t)\equiv
e^{ih_{\B}t}\hat O e^{-ih_{\B}t}.
$

We then turn to the solvent induced and vibration--solvent correlated effects on the EET system.
Similar to \Eq{phixx}, let us define
\begin{align}\label{varphi}\begin{split}
   \varphi_{xy}(t)&\equiv i\la[\hat X^{\B}_{\B}(t),
     \hat Y_{\B}]\ra_{\B}=i\la[\hat Y^{\B}_{\B}(t),
     \hat X_{\B}]\ra_{\B},
\\
   \varphi_{yy}(t)&\equiv i\la[\hat Y^{\B}_{\B}(t),
     \hat Y_{\B}]\ra_{\B},
\end{split}\end{align}
and $\ti\varphi_{xy}(\w)$ and $\ti\varphi_{yy}(\w)$ are defined as \Eq{laplace}.
Note the second identity of \Eq{varphi} exists only in the current model.
It is easy to verify that
\be\label{xxyy0}\begin{split}
 &\ti\varphi_{xx}(0)=\sum_k \frac{c_k^2}{\w_k},\\
 &\ti\varphi_{xy}(0)=\sum_kc_kd_k,\\
 &\ti{\varphi}_{yy}(0)=\sum_k\w_{k}d_k^2.
\end{split}\ee
Thus $\hat F_{1}$ of \Eq{Fvib} can now be recast as
\begin{align}\label{FD0q}
   \hat F_{1}=D_1\hat q,
\end{align}
with
\be\label{barD}
D_1=\ti\varphi_{xy}(0)-\big[\Omega+\ti\varphi_{xx}(0)\big]D.
\ee

We are now in the position to obtain $\Phi_{ab}(t)$ of \Eq{Phidef_sec2} via their frequency resolutions
as long as the solvent response functions $\varphi_{xx}(t)$, $\varphi_{xy}(t)$, and $\varphi_{yy}(t)$ are known.
From \Eq{FD0q}, together with \Eqs{chiqqt} and (\ref{chiqqw}), we have immediately
\be\label{Phivibw}
  \wti\Phi_{11}(\w)= D_1^2\wti\chi_{qq}(\w).
\ee
Following the procedure of the establishment of
the system-bath entanglement theorem in Ref.\ \onlinecite{Du20034102},
we can further obtain
\begin{align}\label{PhinulBw}
\wti\Phi_{12}(\w)&=\wti\Phi_{21}(\w)
\nl &
= D_1^{-1}[D\ti\varphi_{xx}(\w)-\ti\varphi_{xy}(\w)]\wti\Phi_{11}(\w),
\end{align}
and
\begin{align}\label{PhiBw}
 \wti\Phi_{22}(\w)&=D^2\ti\varphi_{xx}(\w)-2D\ti\varphi_{xy}(\w)+\ti{\varphi}_{yy}(\w)
\nl &\quad   +D_1^{-1}[D\ti\varphi_{xx}(\w)-\ti\varphi_{xy}(\w)]\wti\Phi_{12}(\w).
\end{align}
The detailed derivation for \Eqs{PhinulBw} and (\ref{PhiBw}) are given in \App{appa}.
Especially notice \Eq{PhivibB} there that
$\Phi_{21}(t)=\Phi_{12}(t)$ in the current model.
Hence all the involving $\Phi_{ab}(t)=-\Phi_{ba}(-t)$ in \Eq{Jphi} are odd functions,
for $a,b=1,2$.  %   % with $\hat F_1=\hat F_{\nul}$ and $\hat F_2=\hat F_{\B}$.
The odd--function property leads to $J_{ab}(\w)={\rm Im}[\wti\Phi_{ab}(\w)]$.
They can then be readily obtained
via \Eqs{Phivibw}--(\ref{PhiBw}),
together with the expression of $\wti \chi_{qq}(\w)$ [cf.\ \Eq{chiqqw}],
as long as the solvent--space responses $\ti\varphi_{xx}(\w)$, $\ti\varphi_{xy}(\w)$, and $\ti\varphi_{yy}(\w)$ are known.
The resulting $J_{ab}(\w)$, characterizing the vibration--plus--solvent
environment statistics, give the correlations $C_{ab}(t)$ via \Eq{FDT},
which will be plugged in the DEOM construction in \Sec{thsec3}.

 Combining \Eqs{Phivibw}--(\ref{PhiBw}), the overall response function
$\Phi(t)\equiv i\la [\hat F^{\env}(t), \hat F^{\env}(0)]\ra_{\env}$, recast in term of the frequency resolution as
\be
\wti\Phi(\w)=\sum_{ab}\wti\Phi_{ab}(\w),
\ee
is finally obtained as
\begin{align}\label{final}
  \wti\Phi(\w)&=D^2\ti\varphi_{xx}(\w)-2D\ti\varphi_{xy}(\w)+\ti{\varphi}_{yy}(\w)
\nl &\quad +
  [D_1+D\ti\varphi_{xx}(\w)-\ti\varphi_{xy}(\w)]^2\,\wti\chi_{qq}(\w).
\end{align}
Similarly, denote
\be\label{Jwfinal}
  J(\w)={\rm Im}\wti\Phi(\w)=\sum_{ab}J_{ab}(\w).
\ee
The second identity is referred from \Eq{Jphi} and the fact that $\Phi(t)$ is an odd function.
The overall reorganization energy is related to the overall response resolution $\wti\Phi(\w)$ via\cite{Yan05187}
\be\label{lambda3}
 \lambda =\frac{1}{2\pi}\int^{\infty}_{-\infty}\!\!\d\w\,
   \frac{J(\w)}{\w}=\frac{1}{2}\wti\Phi(0)\,.
\ee
It recovers the expression of $\lambda$ in \Eq{lambda} by substituting \Eqs{chiqqw} and (\ref{xxyy0})
into \Eq{final} for $\w=0$.

\section{DEOM formalism}
\label{thsec3}

With the overall environment influence response function [cf.\ \Eq{final}] obtained,
the EET dynamics and spectra can be simulated via the DEOM approach.
DEOM has been applied to study the Herzberg--Teller vibronic coupling
in our previous work.\cite{Zha16204109}
But the correlated non-polarized solvent effects were not included there.
In this section we will give a general DEOM construction in the existence of correlated polarized and non-polarized environments.

\subsection{The system--plus--environment Hamiltonian}

For generality, let us recast the total Hamiltonian of \Eq{Htot0} into the
system--plus--environment form as [cf.\ \Eqs{lambda} and (\ref{F12})]
\be\label{HT0}
  H_{\T}(t) = H_{\exc} +H_{\env}
+\sum_{a=1,2}\hat Q_a\hat F_a-  \hat\mu_{\T}\varepsilon(t),
\ee
with
\be
H_{\exc}=(\Delta+\lambda)\hat{B}^{\dg}\hat{B},\ee
and
\be
\hat Q_1=\hat Q_2=\hat{B}^{\dg}\hat{B}.
\ee
The form of separated $\hat F_1$ and $\hat F_2$ (as well as $\hat Q_1$ and $\hat Q_2$ although they are equal)
has to be adopted for the later DEOM simulation
due to the fact that they may participate differently in the total dipole moment.
In this work the vibration is considered of optical polarization while the solvent not.
The total dipole moment operator assumes the form in the Herzberg--Teller approximation as
\be\label{hat_muB}
\hat \mu_{\T}=\sum_{a=1,2}\hat D_a(u_a+v_a\hat F_a),
\ee
with
\bsube\label{hat_muB_1}
\be \label{hat_muB_11}
\hat D_1=\hat{B}^{\dg}+\hat{B},\ \ u_1=\mu_{\exc},\ \ v_1=\nu_{\nul},
\ee
for the non-Condon vibronic coupling mode,
while
\be\label{hat_muB_22}
\hat D_2=0\quad {\rm and}\quad u_2=v_2=0,
\ee
\esube
for the non-polarized solvent.
%%%
$\mu_{\exc}$ and $\nu_{\nul}$ characterize the excitonic system dipole strength and the non-Condon vibronic coupling strength, respectively.
$\varepsilon(t)$ in \Eq{HT0} is the classical external field.
Generally, DEOM theory deal with arbitrary $H_{\exc}$, $ \{\hat Q_{a}\}$, and $\{\hat D_a\}$.
Influences of $\{\hat F_a\}$ are exerted via their correlation functions $\{C_{ab}(t)\}$ in the environmental space.
The $\hat F_1$ and $\hat F_2$ are associate with the intramolecular vibrational mode
and the solvent degrees of freedom, respectively, as given in \Eqs{Fdecom} with (\ref{XBYB}).
Their participations will be treated in a unified way in the DEOM formalism, see the following subsections,
although the vibrational mode is optically polarizable,
while the solvent not.
Their difference in the optical activity is reflected just via setting the parameters, $u_a$ and $v_a$, cf.\ \Eqs{hat_muB_1}.
Their correlated overall environmental statistical properties have been derived in the previous section.
The key results are give in \Eqs{Phivibw}--(\ref{PhiBw}), in terms of the bath interaction response functions.
The bath correlation functions, on basis of which the DEOM is constructed,
are obtained via the the fluctuation--dissipation theorem, \Eq{FDT}.

\subsection{Bath correlations}

The DEOM construction starts with an exponential expansion of correlation function satisfying \Eq{FDT} as
\be\label{FF_exp0}
  \la \hat F^{\env}_a(t)\hat F^{\env}_b(0)\ra_{\env}
 = \sum_{j} \eta_{abj} e^{-\gamma_{abj} t}.
\ee
This can generally be achieved via certain
sum--over--poles expansion on the Fourier integrand
of \Eq{FDT}, followed by the Cauchy's contour
integration in the low--half plane.
Poles arising
from the Bose function, $1/(1-e^{-\beta\w})$,
are all real, while
those from $J_{ab}(\w)$ are either real
or complex--conjugate paired.
%As inferred from the symmetry properties
%$J^{\ast}_{ab}(\w)=-J_{ab}(-\w)=J_{ba}(\w)$,
Define the associated index $\bar j$ via
$\gamma_{ab\bar j}\equiv \gamma^{\ast}_{abj}$ that
must also be an exponent of \Eq{FF_exp0}.
Due to the time--reversal relation,
$\la \hat F^{\env}_b(0)\hat F^{\env}_a(t)\ra_{\B}
 =\la \hat F^{\env}_a(t)\hat F^{\env}_b(0)\ra^{\ast}_{\B}$,
we have
%%%
\be\label{FF_exp_rev0}
  \la \hat F^{\env}_b(0)\hat F^{\env}_a(t)\ra_{\env}
%=\!\sum_{j=1}^{J} \eta^{\ast}_{abj} e^{-\gamma^{\ast}_{abj} t}
= \sum_{j} \eta^{\ast}_{ab{\bar j}}e^{-\gamma_{abj} t} .
\ee
For convenience in the later DEOM construction,
we recast \Eqs{FF_exp0} and (\ref{FF_exp_rev0}) as
\begin{align}\label{FF_exp}\begin{split}
  \la \hat F^{\env}_a(t)\hat F^{\env}_b(0)\ra_{\env}
 &= \sum_{\kappa} \eta_{ab\kappa} e^{-\gamma_{\kappa} t},
 \\
  \la \hat F^{\env}_b(0)\hat F^{\env}_a(t)\ra_{\env}&= \sum_{\kappa} \eta^{\ast}_{ab\bar\kappa}e^{-\gamma_\kappa t}.
\end{split}
\end{align}
Here, $\gamma_{\kappa}$ runs over all involved exponents in $\{\gamma_{abj}\}$
but with $\eta_{ab\kappa}$ or $\eta^{\ast}_{ab\bar\kappa}$ being
zero if not really among the terms in \Eq{FF_exp0} or \Eq{FF_exp_rev0}.

\subsection{Dissipaton algebra and the construction of DEOM}
\label{thsec2B}

The dissipaton decomposition
on the hybridization environment operator reads\cite{Yan14054105,Yan16110306}
\be\label{F_in_f}
 \hat F^{\env}_a = \sum_{\kappa}\hat f_{a\kappa}.
\ee
This decomposition recovers \Eqs{FF_exp}, by assuming that dissipatons are statistically independent,
with their correlation functions $(t>0)$
\be\label{ff_corr}
\begin{split}
 \la \hat f_{a\kappa}(t)\hat f_{b\kappa'}(0)\ra_{\env}
 &= \delta_{\kappa\kappa'}\eta_{ab\kappa}e^{-\gamma_{\kappa} t},
\\
 \la \hat f_{b\kappa'}(0)\hat f_{a\kappa}(t)\ra_{\env}
 &= \delta_{\kappa\kappa'}\eta^{\ast}_{ab{\bar \kappa}}e^{-\gamma_{\kappa} t}.
\end{split}
\ee
%%%
These lead to the generalized diffusion equation reading
\be\label{diff}
 {\rm tr}_{\env}\Big[\Big(\frac{\partial}{\partial t} \hat f_{a\kappa}\Big)_{\env}\rho_{\T}(t)\Big]
 =-\gamma_{\kappa}\,
    {\rm tr}_{\env}\big[\hat f_{a\kappa}\rho_{\T}(t)\big] .
\ee
%%%
In the DEOM construction below,
this will be used together with the Heisenberg equation--of--motion in the bare environment,
\be\label{hB_Heisenberg}
  \Big(\frac{\partial}{\partial t} \hat f_{a\kappa}\Big)_{\env}=-i[\hat f_{a\kappa},H_{\env}].
\ee

 The dynamical variables in DEOM are called the dissipaton density operators (DDOs),
defined as:
\be\label{DDO}
 \rho^{(n)}_{\textbf{n}}(t)\equiv {\rm tr}_{\env}\Big[
  \Big(\prod_{a\kappa} \hat f^{n_{a\kappa}}_{a\kappa}\Big)^\circ
  \rho_{\T}(t)\Big].
\ee
Here, $n=\sum_{a\kappa} n_{a\kappa}$ and $\textbf{n}=\{n_{a\kappa}\}$
that is an ordered set of the occupation numbers, $n_{a\kappa}=0,1,\cdots$,
on individual dissipatons.
The circled parentheses, $(\cdots)^{\circ}$, is irreducible notation.
For bosonic dissipatons it follows that
$(\hat f_{a\kappa}\hat f_{b\kappa'})^{\circ}=(\hat f_{b\kappa'}\hat f_{a\kappa})^{\circ}$.

The key ingredient in the dissipaton algebra
is the generalized Wick's theorem:
\bsube\label{Wick12}
\begin{align}\label{Wick1}
 &\quad {\rm tr}_{\env}\Big[\Big(\prod_{a\kappa} \hat f^{n_{a\kappa}}_{a\kappa}\Big)^\circ
   \hat f_{b\kappa'} \rho_{\T}(t)\Big]
\nl&=
 \rho^{(n+1)}_{{\bf n}^{+}_{b\kappa'}}(t)+\sum_{a\kappa} n_{a\kappa}\la\hat f_{a\kappa}\hat f_{b\kappa'}\ra^{>}_{\env}
   \rho^{(n-1)}_{{\bf n}^{-}_{a\kappa}}(t),
\end{align}
and
\begin{align}\label{Wick2}
 &\quad {\rm tr}_{\env}\Big[\Big(\prod_{a\kappa} \hat f^{n_{a\kappa}}_{a\kappa}\Big)^\circ
   \rho_{\T}(t)\hat f_{b\kappa'} \Big]
\nl&=
 \rho^{(n+1)}_{{\bf n}^{+}_{b\kappa'}}(t)+\sum_{a\kappa} n_{a\kappa}\la\hat f_{b\kappa'}\hat f_{a\kappa}\ra^{<}_{\env}
   \rho^{(n-1)}_{{\bf n}^{-}_{a\kappa}}(t).
\end{align}
\esube
Here, ${\bf n}^{\pm}_{a\kappa}$ differs from ${\bf n}$ only
at the specified $\hat f_{a\kappa}$-disspaton occupation number,
$n_{a\kappa}$, by $\pm 1$
and
\be\label{ff0}
\begin{split}
 \la\hat f_{a\kappa}\hat f_{b\kappa}\ra^{>}_{\env}
 &\equiv \la\hat f_{a\kappa}(0+)\hat f_{b\kappa'}(0)\ra_{\env} = \eta_{ab\kappa}\delta_{\kappa\kappa'},
\\
 \la\hat f_{b\kappa'}\hat f_{a\kappa}\ra^{<}_{\env}
 &\equiv \la\hat f_{b\kappa'}(0)\hat f_{a\kappa}(0+)\ra_{\env} = \eta^{\ast}_{ab{\bar \kappa}}\delta_{\kappa\kappa'}.
\end{split}
\ee

 The DEOM can now be readily constructed
by applying $\dot{\rho}_{\T}(t)=-i[H_{\T}(t),\rho_{\T}(t)]$,
to the total composite density operator in \Eq{DDO};
i.e.,
\be\label{DDO_dot}
 \dot\rho^{(n)}_{\textbf{n}}(t)= -i\, {\rm tr}_{\env}\Big\{
  \Big(\prod_{a\kappa} \hat f^{n_{a\kappa}}_{a\kappa}\Big)^\circ
  [H_{\T}(t),\rho_{\T}(t)]\Big\}.
\ee
To proceed, let us recast the total Hamiltonian,
\Eq{HT0} with \Eqs{hat_muB} and (\ref{F_in_f}), as
\begin{align}\label{HT}
  H_{\T}(t) =& \Big[H_{\exc}-\sum_a\mu_a\hat D_a\varepsilon(t)\Big] +H_{\env}
\nl &
+\sum_{a\kappa}[\hat Q_{a}-v_{a}\hat D_a\varepsilon(t)]\hat f_{a\kappa}.
\end{align}
Using
\Eq{diff} with \Eq{hB_Heisenberg} for the action of $H_{\env}$,
and \Eqs{Wick12}--(\ref{ff0})
for the action of the last term in \Eq{HT}, we obtain from \Eq{DDO_dot} the DEOM reading
\begin{align}\label{DEOM}
 \dot\rho^{(n)}_{\bf n}&=
 -\Big[i{\cal L}(t)+\sum_{a\kappa} n_{a\kappa}\gamma_\kappa\Big]\rho^{(n)}_{\bf n}
  -i\sum_{a\kappa}\wti{\cal A}_{a}(t)\rho^{(n+1)}_{{\bf n}_{a\kappa}^+}
\nl&\quad
  -i\sum_{a\kappa}n_{a\kappa}\wti{\cal C}_{a\kappa}(t)
   \rho^{(n-1)}_{{\bf n}_{a\kappa}^-} .
\end{align}
Here, the involved superoperators are
\bsube\label{calABC}
\be\label{calA}
{\cal L}(t)\hat O \equiv {\cal L}_{\exc}\hat O-\varepsilon(t)\sum_a u_a{\cal V}_a\hat O,
\ee
and
\begin{align}\label{calB}
\wti{\cal A}_a(t)\hat O&\equiv{\cal A}_a\hat O- \varepsilon(t)v_{a}{\cal V}_a\hat O,
\\ \label{calC}
\wti{\cal C}_{a\kappa}(t)\hat O
&\equiv {\cal C}_{a\kappa}\hat O-\varepsilon(t){\cal D}_{a\kappa}\hat O.
\end{align}
\esube
with the field--free superoperators
\bsube\label{ABCD}
\be
{\cal L}_{\exc}\hat O\equiv [ H_{\exc},\hat O],\  {\cal V}_a\hat O\equiv [\hat D_a,\hat O],\
{\cal A}_{a}\hat O\equiv [ \hat Q_a,\hat O],
\ee
and
\begin{align}
{\cal C}_{a\kappa}\hat O&\equiv \sum_{b}
        (\eta_{ab\kappa} \hat Q_{b}\hat O
        -\eta^{\ast}_{ab\bar \kappa}\hat O\hat Q_{b} ),
\\
 \label{calD}
{\cal D}_{a\kappa}\hat O&\equiv\sum_{b}v_{b}
        (\eta_{ab\kappa}\hat D_b\hat O
        -\eta^{\ast}_{ab\bar \kappa}\hat O\hat D_b).
\end{align}
\esube
Equations (\ref{DEOM})--(\ref{ABCD}) constitute the final DEOM formalism for the hybridized dynamics of
system and its environment, coupled to external fields on Herzberg--Teller polarization [cf.\ \Eq{hat_muB}].

\subsection{Entangled system--and--environment polarization}
\label{specdeom}

In this subsection, we elaborate the procedure to evaluate the entangled system--and--environment polarization
via DEOM.
%The vast number of DDOs, $\{\rho^{(n)}_{\bf n}(t); n=0,1,\cdots\}$, as system--subspace operators, constitute the DDO--space.
Consider the total polarization,
\be\label{PTt_def}
  P_{\T}(t)={\rm Tr}[\hat\mu_{\T}\rho_{\T}(t)] =
    {\rm tr}_{\tS}{\rm tr}_{\env}\big[\hat\mu_{\T}\rho_{\T}(t)\big].
\ee
Here [cf.\ \Eqs{hat_muB} and (\ref{F_in_f})],
\be\label{hat_muT}
  \hat\mu_{\T}=\sum_a \hat D_a \big(u_a+v_a\sum_{\kappa}\hat f_{a\kappa}\big).
\ee
%%%
With \Eq{DDO}, the dissipaton--space evaluation on the Herzberg--Teller polarization can be
expressed as
\be\label{PTt_DDOs}
 P_{\T}(t)=\sum_a{\rm tr}_{\tS}\Big\{\hat D_a \Big[u_a\rho^{(0)}(t)
   +v_a\sum_{\kappa}\rho^{(1)}_{a\kappa}(t)\Big]\Big\}.
\ee
%%%%
The involved $\rho^{(0)}(t)$ and $\{\rho^{(1)}_{a\kappa}(t)\}$
are propagated via \Eq{DEOM} with the dressing field.

 Focusing on the linear absorption spectra,
one can also evaluate the dipole--dipole correlation function in the scenario as
\be\label{dipole_corr}
 %C_{A}(t)\equiv
 \la \hat\mu_{\T}(t)\hat\mu_{\T}(0)\ra
\equiv
  {\rm Tr}(\hat\mu_{\T} e^{-i{\cal L}_{\M}t}\hat\mu_{\T}\rho^{\rm eq}_{\T}).
\ee
Here, ${\cal L}_{\M}\,\cdot\,\equiv [H_{\M},\cdot\,]$
and $\rho^{\rm eq}_{\T}=|0\ra\la 0|e^{-\beta H_{\env}}/{\rm Tr}e^{-\beta H_{\env}}$
are the total composite field--free matter Liouvillian
and the thermal equilibrium density operator, respectively, cf.\ \Eq{HMtot}.
%%%
 The evaluation on the dipole--dipole correlation
function via DEOM goes with the following steps for general cases.
\begin{enumerate}
\item Determine the steady--state correspondence
of $\rho^{\rm eq}_{\T}\Rightarrow
\big\{\rho^{(n)}_{{\bf n};{\rm eq}}\big\}$
evaluated as the solutions to $\dot\rho^{(n)}_{\bf n}=0$ of the field--free \Eq{DEOM}.\cite{Zhe09124508,Din11164107}
For general systems, a self-consistent iteration approach
has been proposed to efficiently solve this problem.\cite{Zha17044105}

\item Identify the correspondence $\hat\mu_{\T}\rho^{\rm eq}_{\T}\Rightarrow
\{\rho^{(n)}_{\bf n}(t=0; \hat \mu_{\T})\}$ by using \Eq{DDO}. We obtain
\begin{align}\label{rhon_t0}
 &\rho^{(n)}_{\bf n}(t=0; \hat \mu_{\T})
\equiv {\rm tr}_{\env}\Big[
  \Big(\prod_{a\kappa} \hat f^{n_{a\kappa}}_{a\kappa}\Big)^\circ
  \big(\hat\mu_{\T}\rho^{\rm eq}_{\T}\big)\Big]
\nl&
 =\sum_a \hat D_a\Big(u_a \rho^{(n)}_{{\bf n};{\rm eq}} +
  v_a  \sum_{\kappa}  \hat \rho^{(n+1)}_{{\bf n}^{+}_{a\kappa};{\rm eq}}\Big)
\nl&\quad
  +\sum_{ab\kappa} v_{b} n_{a\kappa}\eta_{ab\kappa}
    \hat D_b\rho^{(n-1)}_{{\bf n}^{-}_{a\kappa};{\rm eq}}\,.
\end{align}
The second identity is obtained by using \Eq{hat_muT} for $\hat\mu_{\T}$,
followed by the generalized Wick's theorem, \Eq{Wick1} with \Eq{ff0}.
%%%
It can also be viewed as the left--action of dipole operator onto the DDOs
in the DEOM, \Eqs{DEOM}--(\ref{ABCD}).

\item The field--free DEOM propagation is then followed
to obtain $\big\{\rho^{(n)}_{\bf n}(t; \hat \mu_{\T})\big\}$,
the DEOM--space correspondence to $e^{-i{\cal L}_{\M}t}(\hat\mu_{\T}\rho^{\rm eq}_{\T})$.

\item Calculate
\Eq{dipole_corr} in terms of the expectation value like \Eq{PTt_DDOs};
i.e.,
\begin{align}
\label{dipole_corr_DDOs}
  %%\la \hat\mu_{\T}(t)\hat\mu_{\T}(0)\ra
\la \hat\mu_{\T}(t)\hat\mu_{\T}(0)\ra
&=\sum_au_a{\rm tr}_{\tS}\Big[\hat   D_a  \rho^{(0)}(t;\hat \mu_{\T})\Big]
 \nl &\quad
   +\sum_av_a{\rm tr}_{\tS}\Big[\hat D_a  \sum_{\kappa}\rho^{(1)}_{a\kappa}(t;\hat \mu_{\T})\Big].
\end{align}
\end{enumerate}
Finally, the linear absorption spectrum of the total matter is transformed as
\be\label{SA}
  S_{A}(\w)={\rm Re}\int_{0}^{\infty}\!\!\d t\,e^{i\w t}\la \hat\mu_{\T}(t)\hat\mu_{\T}(0)\ra.
\ee

\section{Numerical demonstration}
\label{thsec4}

For the numerical demonstration, we apply Drude model for the
solvent responses $\ti\varphi_{xx}(\w)$, $\ti\varphi_{xy}(\w)$, and $\ti\varphi_{yy}(\w)$, i.e.\
\be\label{Drude}
 \ti\varphi_{cc'}(\w)=\frac{2\eta_{cc'}\gamma}{\gamma-i\w},
\ee
with $c,c'=x,y$.
$\{\eta_{cc'}\}$ are real and non-negative parameters which satisfy
\be \label{schwartz}
\eta_{xy}^2\leq \eta_{xx}\eta_{yy}.
\ee
We introduce further two dimensionless factors to reflect the relations among them:
\begin{align}
g_1\equiv \eta_{xy}/(D\eta_{xx})
\quad \text{and} \quad
g_2\equiv(D\eta_{xy})/\eta_{yy}.
\end{align}
The relation \Eq{schwartz} requires $g_1g_2\leq 1$.
The total transition dipole moments takes the form of \Eq{hat_muB} with \Eq{hat_muB_1}.
Inferred from \Eq{uuulast}, we set
\be\label{mudef}
 \mu_{\nul}\equiv \Omega \nu_{\nul},
\ee
to characterize the strength of the vibrational dipole moment.
We select three types of cases for the following demonstrations.
\begin{enumerate}
\item Brownian--vibration cases: $g_1=g_2=1$. It leads to
$\wti\Phi_{12}(\w)=\wti\Phi_{21}(\w)=\Phi_{22}(\w)=0$ [cf.\ \Eqs{PhinulBw} and (\ref{PhiBw})], whereas
\be
  \wti\Phi_{11}(\w)= \Omega^2D^2\wti\chi_{qq}(\w),
\ee
as inferred from \Eqs{barD} and (\ref{Phivibw}).
Here the vibrational mode in the solvent behaves as a Brownian oscillator.
While the solvent will not directly affect the EET dynamics.

\item  Un-synergetic vibration--solvent cases: $g_1=1$ but $g_2\neq 1$.
This results in $\wti\Phi_{12}(\w)=\wti\Phi_{21}(\w)=0$ and
\bsube
\begin{align}
  \wti\Phi_{11}(\w)&= \Omega^2D^2\wti\chi_{qq}(\w),
  \\
    \wti\Phi_{22}(\w)& =(1-g_2) \ti\varphi_{yy}(\w).
\end{align}
\esube
In the following demonstrations, these conditions will be called ``un-synergetic'' to specifically refer to the cases that
the environmental cross-- response and correlation functions between the Brownian vibrational mode
and the diffusive solvent vanish.

\item General correlated vibration--solvent cases: $g_1\neq 1$ and $g_2\neq 1$.
These are general cases with all the environment influence components,
\Eqs{Phivibw}--(\ref{PhiBw}), being nonzero.
\end{enumerate}

%%\subsection{Linear absorption}

\begin{figure*}
  \includegraphics[width=1.0\textwidth]{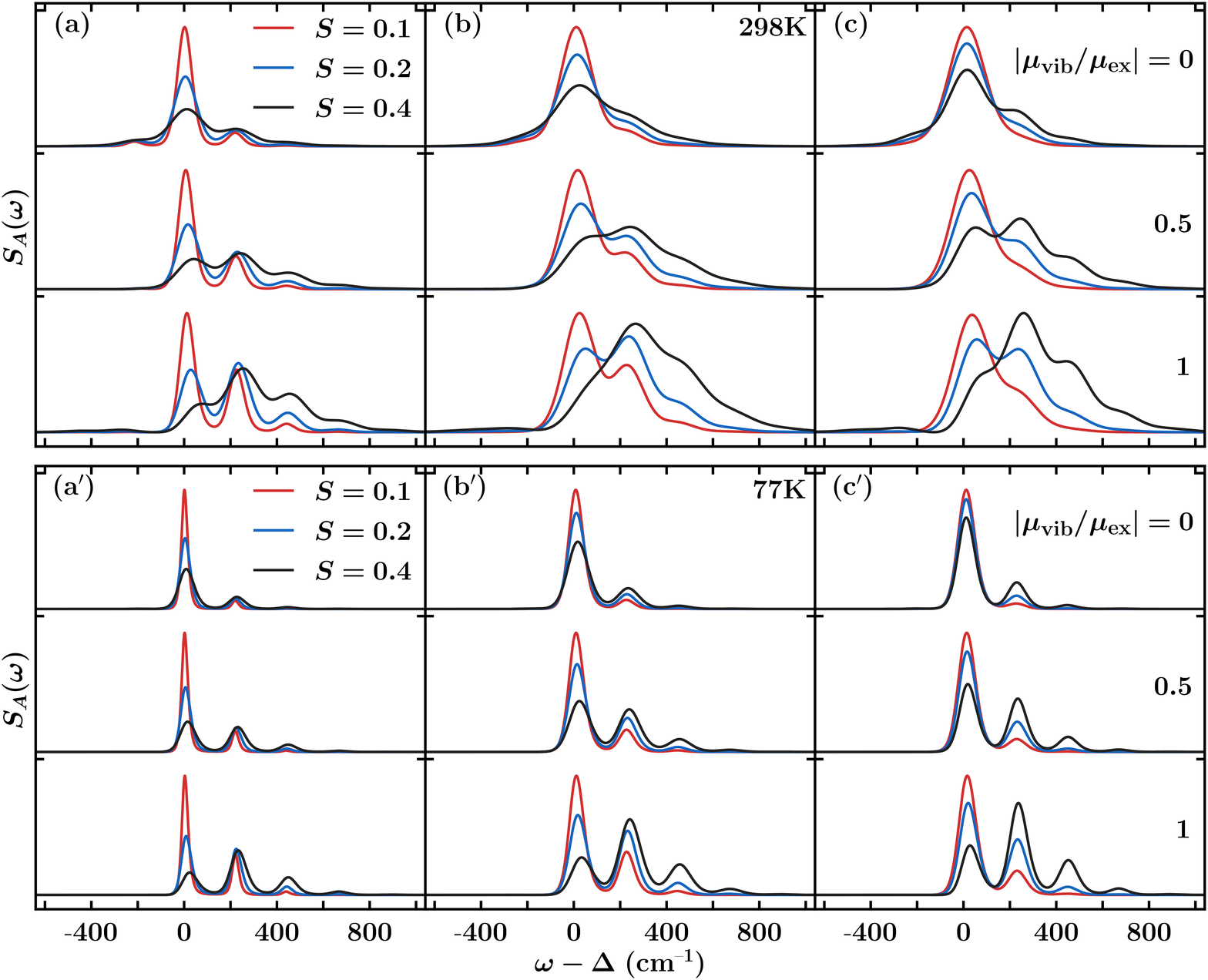}
  \caption{
  Linear absorption lineshapes at the temperatures $T=298$K [upper panels, (a), (b), and (c)] and $T=77$K [lower panels, (a$'$), (b$'$), and (c$'$)],
   in the cases of Brownian--vibration [left panels, (a) and (a$'$)],
    un-synergetic vibration--solvent [middle panels, (b) and (b$'$)],
   and  general correlated vibration--solvent [right panels, (c) and (c$'$)].
   Huang-Rhys factors are selected as $S=0.1,\ 0.2,\ 0.4$ with different vibrational dipole strengths, $|\mu_{\nul}/\mu_{\exc}|=0,\ 0.5,\ 1$.
   For the un-synergetic vibration--solvent case, we select $g_2=0.5$;
   whereas for the general correlated vibration--solvent case, we choose $g_2=g_1^{-1}=\sqrt{2S}$.
   The other parameters are: $\Delta=10000\cm$, $\Omega=200\cm$, and $\gamma= \eta_{xx}=20\cm$.
}
\label{fig1}
\end{figure*}

Analytic solution can be derived for the linear absorption of the monomer system studied in this work.
With the derivation detailed in \App{appb}, the final result is summarized as follows.
\bsube\label{Ana_final}
\begin{align}
\la \hat\mu_{\T}(t)\hat\mu_{\T}(0)\ra&=e^{-i(\Delta+\lambda)t} e^{-g(t)},
\end{align}
with
\begin{align}
e^{-g(t)}=\bigg[&\Big(\mu_{\exc}-i\nu_{\nul}\int_{0}^{t}\!\!{\rm d}\tau \la \hat F^{\env}_{1}(\tau)\hat F^{\env}(0)\ra\Big)
\nl &
\times \Big(\mu_{\exc}-i\nu_{\nul}\int_{0}^{t}\!\!{\rm d}\tau \la \hat F^{\env}(\tau)\hat F^{\env}_{1}(0)\ra\Big)
\nl &
+\nu_{\nul}^2\la \hat F^{\env}_{1}(t)\hat F^{\env}_{1}(0)\ra\bigg]e^{-g_0(t)},
\end{align}
and
\begin{align}
g_0(t)&=\int_{0}^{t}\!\!{\rm d}\tau\!\!\int_{0}^{\tau}\!\!{\rm d}\tau'\,\la \hat F^{\env}(\tau)\hat F^{\env}(\tau')\ra.
\end{align}
\esube
Expressions in the gas--phase condition are further derived in \App{appc}, via two distinct methods.
Given there is also the zero-temperature limit.
The numerical DEOM results, obtained via the procedure described in \Sec{specdeom},
 via DDOs,
have all been confirmed to be consistent with the analytic solutions,
and exhibited in \Fig{fig1}.
 Note that in the present simulations, the step No.1 in \Sec{specdeom} can actually be skipped since
the EET system is initially thermally equilibrated at the ground electronic state.

\begin{figure*}
\includegraphics[width=1.0\textwidth]{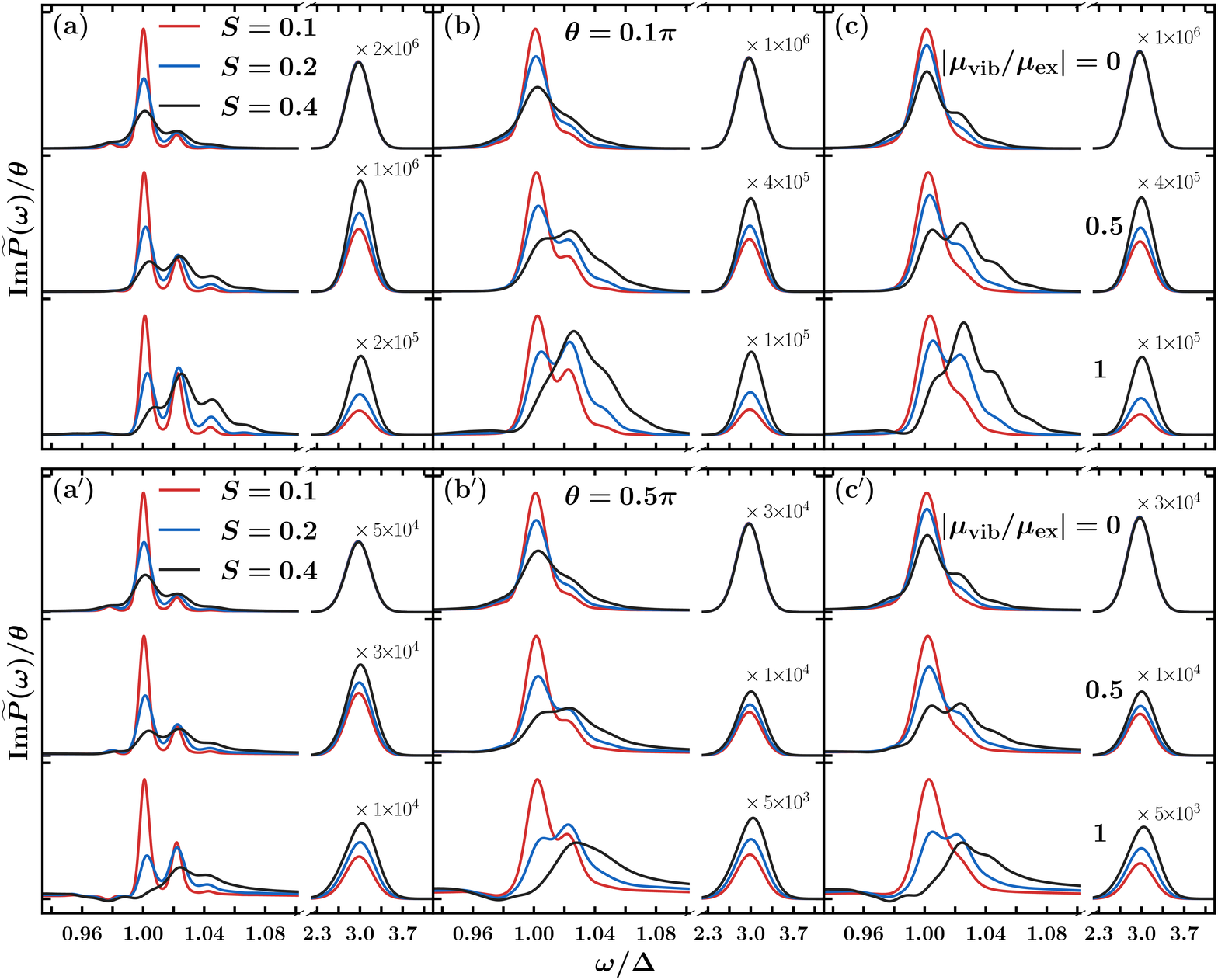}
\caption{Strong field induced polarization spectra (scaled by the flipping angle $\theta$)
         for $\theta=0.1\pi$ [upper panels, (a), (b), and (c)] and $\theta=0.5\pi$  [lower panels, (a$'$), (b$'$), and (c$'$)] at $T=298$K,
          in the cases of Brownian--vibration [left panels, (a) and (a$'$)],
    un-synergetic vibration--solvent [middle panels, (b) and (b$'$)],
   and  general correlated vibration--solvent [right panels, (c) and (c$'$)].
         Other parameters are same as in \Fig{fig1}.
}\label{fig2}
\end{figure*}

Figure \ref{fig1} depicts the monomer absorption spectra in the Brownian--vibration, un-synergetic,
and general correlated vibration--solvent cases at different temperatures,
with various Huang-Rhys factors, $S=D^2/2$, and vibrational dipole strengths $\mu_{\nul}$ [cf.\ \Eq{mudef}].
We select $\Delta=10000\cm$ and $\Omega=200\cm$.
Other parameters are indicated in the figure or given in the caption.
For the polarized vibronic degree of freedom, $\hat F_1$, which behaves as a Browinan oscillator under the influence of solvent,
the effective frequency has been analyzed in Ref.\ \onlinecite{Xu09074107} to be
\be
\Omega_{\rm eff}^2=\lim_{t\rightarrow \infty}\frac{\ddot\chi_{qq}^2(t)-\dddot\chi_{qq}(t)\dot\chi_{qq}(t)}{\dot\chi_{qq}^2(t)-\ddot\chi_{qq}(t)\chi_{qq}(t)},
\ee
with $\chi_{qq}(t)$ being defined in \Eq{chiqqt}.
It characterizes the spectroscopic features in \Fig{fig1}, rather than the original frequency $\Omega$.
Based on the selected solvent parameters, the $\Omega_{\rm eff}$ is obtained to be $1.095\Omega$.
We can see that the solvent effect is more important at high temperature than at low temperature,
by comparison to the corresponding gas-phase results, depicted in \Fig{fig3} in \App{appc}.
The solvent broadening is observed by comparing (b) to (a)  and
the coherence enhancement, due to the synergetic vibration--solvent correlation, is observed by comparing (c) to (b).

Figure \ref{fig2} shows the strong field induced polarization spectra,
\be\label{freq-polar}
  \ti{P}(\omega)\equiv \int_{-\infty}^{\infty}\!\!\! \d t\,
   e^{i\omega t}\left[P_{\T}(t)-P_{\T}^{\rm eq}\right],
\ee
evaluated via the DEOM simulations.
$P_{\T}^{\rm eq}$ is the equilibrium polarization of the total system.
Right-hand-sides of the break exhibit the nonlinear triple-frequency signals.
The external field is adopted as a Gaussian pulse, with the envelope being
\be\label{Gauss-p}
  \mu_{\exc}\varepsilon(t)=\frac{\theta}{\sqrt{2\pi}\sigma}
  \exp\left(-\frac{t^2}{2\sigma^2}\right)\cos(\omega_0 t).
\ee
Here, $\omega_0$ is the carrier frequency.
$\theta$ and $\sigma$ denotes the pulse strength and width. We set $\omega_0=10040\cm$ and $\sigma=4$ fs.
$\theta$ is chosen as $0.1\pi$ (upper panels) and $0.5\pi$ (lower panels),
for the relative weak and strong dressing fields, respectively.
Both the nonlinear polarization and the dressed effect in the linear regime
vanish in the weak field limit, as indicated in the upper panels.
Similar to \Fig{fig1}, the (b) and (b$'$) panels in
\Fig{fig2} exhibit the solvent broadening effect in comparison to the (a) and (a$'$) panels;
whereas the (c) and (c$'$) panels
show the  synergetic vibration--solvent correlation induced coherence enhancement
in comparison to the (b) and (b$'$) panels.

%\subsection{Strong--field polarization}

\section{Summary}
\label{thsec5}

To summarize, this work studies the correlated effect between the non-Condon vibronic coupling
and the surrounding solvent influences.
We start from a physical monomer model which involves
 correlated exciton--vibration, exciton--solvent, and
vibration--solvent interactions, with both the exciton and the vibrational mode being optical polarizable.
The overall statistical vibration--plus--solvent environmental effects are analyzed detailedly in obtaining
the overall environmental interaction response functions.
On basis of them, we derive the analytic linear absorption solutions and construct
the general dissipaton--equation--of--motion (DEOM) formalism to carry out simulations on nonlinear spectroscopies and arbitrary systems.
Numerical demonstrations in either the linear absorption or strong field regime
clearly show the coherence enhancement due to the synergetic vibration--solvent correlation.
This observed feature is expected to be detected in the multi-exciton systems and multi-dimensional spectroscopies as well.
The solvent-polarization induced Fano interference is also to be considered in future work.

\vspace{0.5 em}
\noindent{\bf Data Availability}:
The data that support the findings of this study are available from the corresponding author upon reasonable request.

\begin{acknowledgments}
Support from the Ministry of Science and Technology of China No.\ 2017YFA0204904,
the National Natural Science Foundation of China No.\ 21633006,
and Anhui Initiative in Quantum Information Technologies
is gratefully acknowledged.
\end{acknowledgments}

\appendix
\section{Derivation for \Eqs{PhinulBw}--(\ref{PhiBw})}
\label{appa}
Note that in the present microscopic model,
\bsube
\begin{align}
  \varphi_{xx}(t)&=\sum_kc^2_k\sin(\w_kt),\\
  \varphi_{xy}(t)&=\sum_k\w_kc_kd_k\sin(\w_kt),\\
  \varphi_{yy}(t)&=\sum_k\w^2_kd^2_k\sin(\w_kt).
\end{align}
\esube
From \Eq{ope_env} together with \Eq{Henv}, we have
\be\label{eq05}
\ddot{\hat x}^{\env}_k(t)=-\omega^2_k\hat x^{\env}_k(t)+\omega_kc_k\hat q^{\env}(t).
\ee
The formal solution to $\hat x^{\env}_k(t)$ is
\begin{align}
   \hat x^{\env}_k(t)&=\hat x_k\cos(\omega_kt)
                      +\hat p_k\sin(\omega_kt)
\nl & \quad +c_{k}
   \int_0^t\!\!{\rm d}\tau\,\sin[\omega_k(t-\tau)]\hat q^{\env}(\tau).
\end{align}
Together with \Eq{XBYB}, the above equation leads to
\bsube\label{XYenv}
\begin{align}
     \hat X^{\env}_{\B}(t)&=\hat X^{\B}_{\B}(t)+\int_0^t\!\!{\d}\tau\,\varphi_{xx}(t-\tau)\hat q^{\env}(\tau),
\\
     \hat Y^{\env}_{\B}(t)&=\hat Y^{\B}_{\B}(t)+\int_0^t\!\!{\d}\tau\,\varphi_{xy}(t-\tau)\hat q^{\env}(\tau).
\end{align}
\esube
As $[\hat X^{\B}_{\B}(t),\hat O]=[\hat Y^{\B}_{\B}(t),\hat O]=0$ for arbitrary vibrational operator $\hat O$,
we obtain [cf.\ \Eqs{Fdecom} and (\ref{FD0q})]
\begin{align}\label{PhivibB}
  \Phi_{21}(t)&= D_0^{-1}\!\!\int_0^t\!\!{\d}\tau\,[D\varphi_{xx}(t-\tau)-\varphi_{xy}(t-\tau)]\Phi_{11}(\tau)
\nl &= \Phi_{12}(t).
\end{align}
The second identity is obtained by $\Phi_{12}(t)=-\Phi_{21}(-t)$, together with the  fact that
$\Phi_{\nul}(t)$, $\varphi_{xx}(t)$, $\varphi_{xy}(t)$ are all odd functions.
Furthermore, \Eqs{XYenv} together with \Eq{FBXY} result in
\begin{align}\label{PhiB}
 \Phi_{22}(t)&=D_0^{-1}\int_0^t\!\!{\d}\tau\,[D\varphi_{xx}(t-\tau)-\varphi_{xy}(t-\tau)]\Phi_{12}(\tau)
 \nl
 &\quad+D^2\varphi_{xx}(t)-2D\varphi_{xy}(t)+\varphi_{yy}(t).
\end{align}
In terms of frequency resolution, \Eq{PhivibB} and \Eq{PhiB} lead to
\Eq{PhinulBw} and
\Eq{PhiBw}, respectively.

\section{Derivation of \Eq{Ana_final}}
\label{appb}

This appendix derives the analytic linear absorption lineshape
in case of non-Condon polarization. % without solvent effects.
%The resulted expression is also helpful for the analysis in the presence of solvent.
For the monomer system in this work, the dipole--dipole correlation, \Eq{dipole_corr},
can be recast as [cf.\ \Eq{hat_muB} with \Eq{hat_muB_1}]
\begin{align}
\la \hat\mu_{\T}(t)\hat\mu_{\T}(0)\ra   %%  \la \hat\mu_{\T}(t)\hat\mu_{\T}(0)\ra
&=e^{-i\Delta t}\big\la e^{iH_{\env}t}(\mu_{\exc}\!+\!\nu_{\nul}\hat F_{1})e^{-iH'_{\env}t}
\nl & \qquad\qquad \times  (\mu_{\exc}\!+\!\nu_{\nul}\hat F_{1})\big\ra.
\end{align}
By applying [cf.\ \Eq{force}]
\be
e^{-iH'_{\env}t}=e^{-i\lambda t}e^{-iH_{\env}t}\exp_{+}\bigg[-i\!\int_{0}^{t}\!\!{\rm d}\tau\,\hat F^{\env}(\tau)\bigg],
\ee
%where the subscript ``$+$'' denotes the time ordering,
we obtain
\begin{align} \label{ephi}
\la \hat\mu_{\T}(t)\hat\mu_{\T}(0)\ra&=e^{-i(\Delta+\lambda)t} e^{-g(t)},
\end{align}
where
\be \label{emgt}
e^{-g(t)}\equiv\bigg\la \hat D'(t)
\exp_{+}\bigg[-i\!\int_{0}^{t}\!\!{\rm d}\tau\,\hat F^{\env}(\tau)\bigg]
\hat D'(0)\bigg\ra,
\ee
with
\be
\hat D'(t)\equiv\mu_{\exc}+\nu_{\nul}\hat F^{\env}_{1}(t).
\ee

To proceed, we introduce the generating function
\begin{align}\label{genfun}
G(t;t',t'')\equiv\,&\bigg \la \exp_{+}\bigg[-i\!\int_{t}^{t''}\!\!{\rm d}\tau\,\hat F_1^{\env}(\tau)\bigg]
\nl &
\times
\exp_{+}\bigg[-i\!\int_{0}^{t}\!{\rm d}\tau \,\hat F^{\env}(\tau)\bigg]
\nl &
\times
\exp_{+}\bigg[-i\!\int_{t'}^{0}\!{\rm d}\tau \,\hat F_1^{\env}(\tau)\bigg]\bigg\ra.
\end{align}
Equation (\ref{emgt}) can be evaluated via
\be\label{emgt1}
e^{-g(t)}=\lim_{\substack{t''\rightarrow t\\ t'\rightarrow 0}} \big(\mu_{\exc}+i\nu_{\nul}\partial_{t''}\big)\big(\mu_{\exc}-i\nu_{\nul}\partial_{t'}\big)G(t;t',t'').
\ee
On the other hand, the Gauss--Wick's theorem leads to \Eq{genfun} the expression
%Due to the Gaussian properties, the generating function can be obtained exactly via the second--order cumulant expansion, reading
\be
G(t;t',t'')=e^{-g_0(t)}e^{-g'(t;t',t'')},
\ee
with
\begin{align}\label{ephi0}
g_0(t)&=\int_{0}^{t}\!\!{\rm d}\tau\!\!\int_{0}^{\tau}\!\!{\rm d}\tau'\,\la \hat F^{\env}(\tau)\hat F^{\env}(\tau')\ra,
\end{align}
and
%\begin{widetext}
\begin{align}
g'(t;t',t'')&=\int_{t}^{t''}\!\!\!{\rm d}\tau\!\!\int_{t}^{\tau}\!\!{\rm d}\tau'\la \hat F_1^{\env}(\tau)\hat F_1^{\env}(\tau')\ra
\nl &\quad
+\int_{t}^{t''}\!\!\!{\rm d}\tau\!\!\int_{0}^{t}\!\!\d\tau'\la \hat F_1^{\env}(\tau)\hat F^{\env}(\tau')\ra
\nl &\quad
+\int_{t}^{t''}\!\!\!{\rm d}\tau\!\!\int_{t'}^{0}\!\!\d\tau'\la \hat F_1^{\env}(\tau)\hat F_1^{\env}(\tau')\ra
\nl &\quad
+\int_{0}^{t}\!\!{\rm d}\tau\!\!\int_{t'}^{0}\!\!\d\tau'\la \hat F^{\env}(\tau)\hat F_1^{\env}(\tau')\ra
\nl &\quad
+\int_{t'}^{0}\!\!{\rm d}\tau\!\!\int_{t'}^{\tau}\!\!{\rm d}\tau'\la \hat F_1^{\env}(\tau)\hat F_1^{\env}(\tau')\ra.
\end{align}
Evidently, only the middle three terms make contributions to \Eq{emgt1}.
We obtain
\begin{align}
\label{ephib3}
e^{-g(t)}=\bigg[&\Big(\mu_{\exc}-i\nu_{\nul}\int_{0}^{t}\!\!{\rm d}\tau \la \hat F^{\env}_{1}(\tau)\hat F^{\env}(0)\ra\Big)
\nl &
\times \Big(\mu_{\exc}-i\nu_{\nul}\int_{0}^{t}\!\!{\rm d}\tau \la \hat F^{\env}(\tau)\hat F^{\env}_{1}(0)\ra\Big)
\nl &
+\nu_{\nul}^2\la \hat F^{\env}_{1}(t)\hat F^{\env}_{1}(0)\ra\bigg]e^{-g_0(t)}.
\end{align}
Equation ({\ref{Ana_final}) is then resulted by substituting \Eq{ephib3}
into \Eq{ephi} together with \Eq{ephi0}.

\section{Gas--phase limit}
\label{appc}

 This appendix considers the gas--phase limit in the absence of solvent.
As the limiting result of \Eqs{SA} with (\ref{Ana_final}), the spectrum can be derived readily and will be given
in the latter half of this appendix. For just the case of gas phase,
the frequency domain Fermi's golden rule will be more convenient.
 That is
\be \label{c1}
S_{A}(\w)=\pi \sum_{m=0}^\infty\sum_{n=0}^\infty P_{n}|\la \psi'_m | \mu_{10}(\hat q)|\psi_n\ra|^2\delta(\w-\w_{m n}),
\ee
with $
\w_{mn}\equiv \Delta+ (m-n)\Omega
$ being  the transition frequency.
Involved  are the vibronic states, $|\psi_n\ra$ and $|\psi'_m\ra$, from $H_{\env}=\frac{1}{2}\Omega(\hat p^2+\hat q^2)$ and $H'_{\env}=\frac{1}{2}\Omega[\hat p^2+(\hat q-D)^2]$, respectively.
The initial equilibrium population of $|n\ra$ is
\be
P_n= (1-e^{-\beta\Omega})e^{- n\beta\Omega}.
\ee
 The electronic  transition dipole moment remains an operator in nuclear subspace, and is given by
\be
\mu_{10}(\hat q)\equiv \la 1|\hat \mu_{\T}|0\ra=\mu_{\exc}+\nu_{\nul}\hat F=\mu_{\exc}-\nu_{\nul}\Omega D\hat q,
\ee
since $\hat F=\hat F_{1}=H'_{\env}-H_{\env}-\lambda_{0}=-\Omega D\hat q$
in the gas--phase limit.

The Franck--Condon factor can be obtained as follows. By exploring the displacement operator, we obtain
\begin{align}\label{fc4}
\la \psi'_m | \mu_{10}(\hat q)|\psi_n\ra&=\la \psi_m |e^{i\hat p D} \mu_{10}(\hat q)|\psi_n\ra
\nl &
=\la \psi_m |e^{-\frac{D}{\sqrt{2}}(\hat a^{\dg}-\hat a)} \mu_{10}(\hat q)|\psi_n\ra
\nl &
=e^{-\frac{D^2}{4}}(\mu_{\exc} \Lambda_{mn}-\nu_{\nul}\Omega\ti\Lambda_{mn})
\end{align}
where
\begin{align}\label{fc5}
\Lambda_{mn}&=\la \psi_m |e^{-\frac{D}{\sqrt{2}}\hat a^{\dg}}e^{\frac{D}{\sqrt{2}}\hat a} |\psi_n\ra,
\nl &
=\sum_{jk}\frac{(-1)^j}{j!k!}\Big(\frac{D}{\sqrt{2}}\Big)^{j+k}\la \psi_m |(a^{\dg})^j\hat a^k |\psi_n\ra
\nl &
=\sum_{jk}\frac{(-1)^j}{j!k!}\Big(\frac{D}{\sqrt{2}}\Big)^{j+k}\!\!\sqrt{\frac{n!m!}{(n-k)!(m-j)!}}\delta_{m-j,n-k}
%\nl &
%%=\sum_{k=0}^{n}\frac{(-1)^{(m-n+k)}}{k!}\sqrt{\frac{n!}{m!}}\binom{m}{n-k}\Big(\frac{D}{\sqrt{2}}\Big)^{m-n+2k},
\nl &
=\sum_{j=0}^{m}\frac{(-1)^{j}}{j!}\sqrt{\frac{m!}{n!}}\binom{n}{m-j}\Big(\frac{D}{\sqrt{2}}\Big)^{n-m+2j},
\nl &
\equiv \sum_{j=0}^{m} K_{m,n,j}\Big(\frac{D}{\sqrt{2}}\Big)^{n-m+2j},
\end{align}
with the convention $\binom{n}{l>n}=0$, and
\begin{align}\label{fc6}
\ti\Lambda_{mn}&=D\la \psi_m |e^{-\frac{D}{\sqrt{2}}\hat a^{\dg}}e^{\frac{D}{\sqrt{2}}\hat a} \hat q|\psi_n\ra,
\nl &
=\frac{D}{\sqrt{2}}(\sqrt{n}\Lambda_{m,n-1}+\sqrt{n+1}\Lambda_{m,n+1}).
\nl &
=\sqrt{n}\sum_{j=0}^{m} K_{m,n-1,j}\Big(\frac{D}{\sqrt{2}}\Big)^{n-m+2j}
\nl &\quad
+S\sqrt{n+1}\sum_{j=0}^{m} K_{m,n+1,j}\Big(\frac{D}{\sqrt{2}}\Big)^{n-m+2j}
\nl &
=\sum_{j=0}^{m}\wti K_{m,n,j}\Big(\frac{D}{\sqrt{2}}\Big)^{n-m+2j},
\end{align}
with $\wti K_{m,n,j}\equiv\sqrt{n}K_{m,n-1,j}
+S\sqrt{n+1}K_{m,n+1,j}$.
Substituting \Eqs{fc5} and (\ref{fc6}) into \Eq{fc4} and denoting
\be\label{Cadd1}
R_{m,n,j}\equiv\mu_{\exc}K_{m,n,j}-\nu_{\nul}\Omega\wti K_{m,n,j},
\ee
 we have
\begin{align}\label{fc7}
\la \psi'_m | \mu_{10}(\hat q)|\psi_n\ra
=e^{-\frac{S}{2}}\sum_{j=0}^{m}R_{m,n,j}\Big(\frac{D}{\sqrt{2}}\Big)^{n-m+2j}.
\end{align}
Substituting it into \Eq{c1} and noticing that it is real,
we obtain
\begin{align} \label{fc8}
S_{A}(\w)=\pi e^{-S}\sum_{m=0}^\infty&\!\sum_{n=0}^\infty\sum_{j=0}^{m}\sum_{k=0}^{m}R_{m,n,j}R_{m,n,k}
\nl & \times
S^{n-m+j+k} P_{n}
\delta(\w-\w_{m n}).
\end{align}

To finalize, we shall convert the summation index in \Eq{fc8} as
\be
\sum_{m=0}^{\infty}\sum_{n=0}^{\infty}=\sum_{M=0}^\infty\sum_{m=M}^{\infty}+\sum_{M=-\infty}^{-1}\sum_{n=-M}^{\infty}\,,
\ee
by denoting $M\equiv m-n$.
We can thus recast \Eq{fc8} as
\begin{align}\label{fc10}
S_{A}(\w)&=\pi \!\!\sum_{M=-\infty}^{\infty}
                        \!\!A_M\delta(\w-\w_{M}),
\end{align}
with $\w_{M}\equiv \w_{mn}$ and for $M\geq 0$
\bsube\label{c11s}
\begin{align} \label{c11a}
A_M= e^{-S}\!\!\sum_{m=M}^{\infty}\!\! P_{m-M}&\!\sum_{j=0}^{m}\sum_{k=0}^{m}R_{m,m-M,j}
\nl &\times R_{m,n-M,k}
S^{j+k-M},
\end{align}
while for $M<0$
\begin{align} \label{c11b}
A_M= e^{-S}\!\!\sum_{n=-M}^{\infty}\!\! P_{n}&\!\sum_{j=0}^{n+M}\sum_{k=0}^{n+M}R_{n+M,n,j}
\nl &\times R_{n+M,n,k}
S^{j+k-M}.
\end{align}
\esube
 Equations (\ref{fc10}) with (\ref{c11s}) and (\ref{Cadd1})
consititute the final frequency domain results on the gas-phase condition.
Particularly, in the zero-temperature limit where $n=0$ and $P_0=1$,
we have
\bsube
\be
\Lambda_{m0}=\frac{(-1)^m}{\sqrt{m!}}\Big(\frac{D}{\sqrt{2}}\Big)^{m},
\ee
\be
\ti\Lambda_{m0}=\frac{(-1)^m}{\sqrt{m!}}\Big(\frac{D}{\sqrt{2}}\Big)^{m}(S-m)=(S-m)\Lambda_{m0},
\ee
\esube
and
\begin{align}\label{fc012}
S_{A}(\w)=\pi e^{-S}\sum_{m=0}^\infty\frac{S^m}{m!}&[\mu_{\exc}-\nu_{\nul}\Omega(S-m)]^2
\nl & \times \delta(\w-\Delta-m\Omega).
\end{align}

\begin{figure*}
  \includegraphics[width=0.618\textwidth]{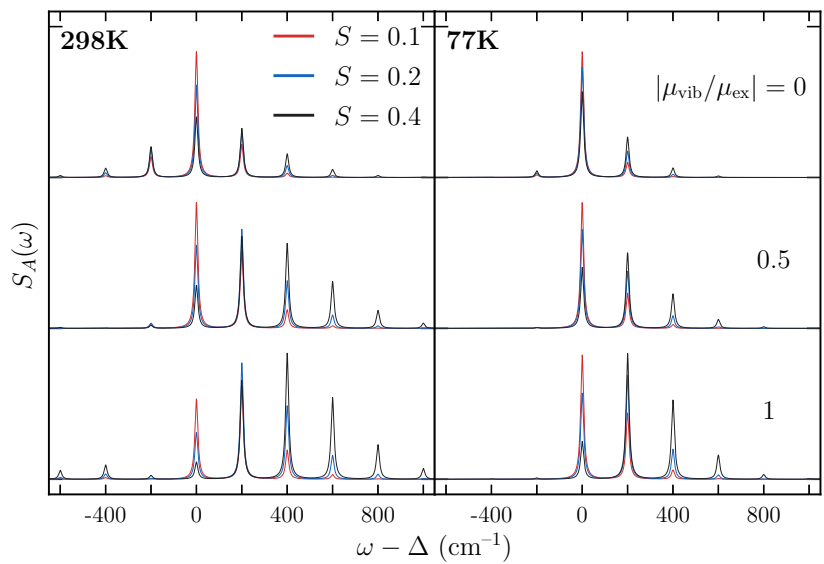}
  \caption{
  Gas--phase linear absorption lineshapes at the temperatures $T=298$K (left) and $T=77$K (right). Other involved parameters are same as in \Fig{fig1}.
}\label{fig3}
\end{figure*}

Alternatively, let us recall \Eqs{SA} with (\ref{Ana_final}) in the gas--phase condition where $\hat F=\hat F_1$. We have
\be \label{SAA}
S_{A}(\w)={\rm Re}\int_{0}^{\infty}\!\!\d t\,e^{i(\w-\Delta-\lambda_0)t}e^{-g(t)},
\ee
with
\begin{align}
\label{ephi3}
e^{-g(t)}=\bigg[&\Big(\mu_{\exc}-i\nu_{\nul}\int_{0}^{t}\!\!{\rm d}\tau \la \hat F^{\env}_{1}(\tau)\hat F^{\env}_1\ra\Big)^2
\nl &
+\nu_{\nul}^2\la \hat F^{\env}_{1}(t)\hat F^{\env}_{1}\ra\bigg]e^{-g_0(t)},
\end{align}
and
\be\label{Cg0}
e^{-g_0(t)}=\exp\bigg[-\int_{0}^{t}\!\!{\rm d}\tau\!\!\int_{0}^{\tau}\!\!{\rm d}\tau'\,\la \hat F_1^{\env}(\tau)\hat F_1^{\env}(\tau')\ra\bigg].
\ee
For the gas--phase model, it is easy to obtain that
\be\label{Cff2}
\la \hat F^{\env}_{1}(t)\hat F^{\env}_{1}\ra_{\env}=\Omega^2 S[(\bar n+1)e^{-i\Omega t}+\bar ne^{i\Omega t}],
\ee
with $\bar n=1/(e^{\beta\Omega}-1)$.
Thus
\be\label{0k2}
\int_{0}^{t}\!\!{\rm d}\tau \la \hat F^{\env}_{1}(\tau)\hat F^{\env}_{1}\ra_{\env}
=-i\Omega S[1-(\bar n+1)e^{-i\Omega t}
 +\bar ne^{i\Omega t}],
\ee
and
\begin{align}\label{0k3}
  \int_{0}^{t}\!\!{\rm d}\tau\!
  \int_{0}^{\tau}\!\!{\rm d}\tau'
&  \la \hat F^{\env}_{1}(\tau')\hat F^{\env}_{1}\ra_{\env}
   =-i\Omega S t+S(2\bar n+1)
\nl &
  -S[(\bar n+1)e^{-i\Omega t}
 +\bar n e^{i\Omega t}].
\end{align}

Equation (\ref{Cg0}) thus results in (noting that $\lambda_0=\Omega S$)
\begin{align}
 e^{-g_0(t)}=e^{i\lambda_0 t} W(t),
\end{align}
with
\begin{align}
W(t)&\equiv e^{-S(2\bar n+1)}e^{S[(\bar n+1)e^{-i\Omega t}
 +\bar n e^{i\Omega t}]}
\nl &=
 \sum_{m=0}^{\infty}\sum_{n=0}^{\infty}e^{-S(2\bar n+1)}\frac{S^{m+n}(\bar n+1)^m\bar n^n}{m!n!}e^{-i(m-n)\Omega t}
\nl &
 \equiv\sum_{m=0}^{\infty}\sum_{n=0}^{\infty}W_{mn}e^{-i(m-n)\Omega t}.
\end{align}
Together with \Eqs{Cff2} and (\ref{0k2}), we obtain from \Eqs{SAA} and (\ref{ephi3}) that
\begin{align}\label{uuu2}
S_A(\w)&=\pi\sum_{m=0}^{\infty}\sum_{n=0}^{\infty}\sum_{k=-2}^2W_{mn}Y_k \delta(\w-\w_{mn}-k\Omega),
\end{align}
with
\bsube
\begin{align}
Y_{2}&=[\nu_{\nul}\Omega S (\bar n+1)]^2,
\\
Y_{1}&=[2\mu_{\exc}+(1-2S)\nu_{\nul}\Omega]\nu_{\nul}\Omega S(\bar n+1),
\\
Y_{0}&=(\mu_{\exc}-\nu_{\nul}\Omega S)^2-2\nu_{\nul}^2\Omega^2S^2\bar n(\bar n+1),
\\
Y_{-1}&=[-2\mu_{\exc}+(1+2S)\nu_{\nul}\Omega]\nu_{\nul}\Omega S\bar n,
\\
Y_{-2}&=(\nu_{\nul}\Omega S \bar n)^2.
\end{align}
\esube
After some rearrangement of summation indices,
\Eq{uuu2}) can be recast in the form of
\begin{align}\label{ffc24}
S_{A}(\w)=\pi\!\!\sum_{M=-\infty}^{\infty}\!\!A'_{M}\delta(\w-\w_{M}).
\end{align}
The details are a bit tedious and not given here.
The equivalence between the $A'_{M}$ in \Eq{ffc24} and $A_M$ of \Eq{c11s}
has been numerically confirmed.
Depicted in \Fig{fig3} are the linear absorption lineshapes in the gas--phase condition,
evaluated via \Eq{ffc24} or \Eq{fc10}, where the $\delta$--functions have been broadened as Lorentz functions.

We consider again the zero-temperature limit where $\bar n=0$, leading to
\begin{align}
W(t)=e^{-S}
 \sum_{m=0}^{\infty}\frac{S^{m}}{m!}e^{-im\Omega t},
\end{align}
and
\bsube
\begin{align}
&Y_{2}=(\nu_{\nul}\Omega S)^2,
\\
&Y_{1}=[2\mu_{\exc}+(1-2S)\nu_{\nul}\Omega]\nu_{\nul}\Omega S,
\\
&Y_{0}=(\mu_{\exc}-\nu_{\nul}\Omega S)^2,
\\
&Y_{-1}=Y_{-2}=0.
\end{align}
\esube
We can then obtain from \Eq{uuu2} that
\begin{align}\label{uuulast}
&S_A(\w)=\pi e^{-S}\!\!\sum_{m=0}^{\infty}\!\!
  \frac{S^{m}}{m!}
  \Big\{(\nu_{\nul}\Omega S)^2\delta(\w-\!\Delta\!-\!(m+2)\Omega)
\nl &\quad +
  [2\mu_{\exc}+(1-2S)\nu_{\nul}\Omega]\nu_{\nul}\Omega S
                                 \delta(\w-\!\Delta\!-\!(m+1)\Omega)
\nl &\quad +
   (\mu_{\exc}-\nu_{\nul}\Omega S)^2
                                 \delta(\w-\!\Delta\!-\!m\Omega)
  \Big\}
\nl
 &=\pi e^{-S}\!\!\sum_{m=0}^{\infty}\!\!
  \frac{S^{m}}{m!}
  \Big\{m(m-1)(\nu_{\nul}\Omega)^2+
   (\mu_{\exc}-\nu_{\nul}\Omega S)^2
\nl &\quad +
  m[2\mu_{\exc}+(1-2S)\nu_{\nul}\Omega]\nu_{\nul}\Omega
 \Big\}
                                 \delta(\w-\!\Delta\!-\!m\Omega)
\nl
 &=\pi e^{-S}\!\!\sum_{m=0}^{\infty}\!\!
  \frac{S^{m}}{m!}
   (\mu_{\exc}\!-\!\nu_{\nul}\Omega S\!+\!m\nu_{\nul}\Omega)^2
                                 \delta(\w-\!\Delta\!-\!m\Omega).
\end{align}
This is equivalent to \Eq{fc012}.

%\bibliographystyle{aiptit}
%\bibliography{bibrefs}

\end{document}